\documentclass[12pt,a4paper]{article}
\title{On Fermionic T-duality of Sigma modes on AdS backgrounds}

\author{Chen-guang Hao, Bin Chen and Xing-chang Song\\ \\
Department of Physics,\\
and State Key Laboratory of Nuclear Physics and Technology,\\
Peking University, Beijing 100871, P.R.
China\footnote{Emails:haocch@126.com,
bchen01@pku.edu.cn,songxc@pku.edu.cn}}
\date{}

\begin{document}
\maketitle
\begin{abstract}
We study the fermionic T-duality symmetry of integrable
Green-Schwarz sigma models on AdS backgrounds. We show that the
sigma model on $AdS_5\times S^1$ background is self-dual under
fermionic T-duality. We also construct new integrable sigma models
on $AdS_2\times CP^n$. These backgrounds could be realized as
supercosets of SU supergroups for arbitrary $n$, but could also be
realized as supercosets of OSp supergroups for $n=1,3$. We find
that the supercosets based on SU supergroups are self-dual under
fermionic T-duality, while the supercosets based on OSp
supergroups are not. However, the reasons of OSp supercosets being
not self-dual under fermionic T-duality are different. For
$OSp(6|2)$ case, corresponding to $AdS_2\times CP^3$ background,
the failure is due to the singular  fermionic quadratic terms,
just like $AdS_4\times CP^3$ case. For $OSp(3|2)$ case, the
failure is due to the shortage of right number of
$\kappa$-symmetry to gauge away the fermionic degrees of freedom,
even though the fermionic quadratic term is not singular any more.
More general, for the supercosets of the OSp supergroups with
superalgebra $B(n,m)$, including $AdS_2\times S^{2n}$ and
$AdS_4\times S^{2n}$ backgrounds,  the sigma models are not
self-dual under fermionic T-duality as well, obstructed by the
$\kappa$-symmetry.
\end{abstract}

\section{Introduction}

Recently, it was found that $\mathcal{N}=4$ SYM  gluon scattering
amplitudes display a non-trivial symmetry called the dual
conformal
invariance\cite{Alday:2007hr,Drummond:2006rz,Drummond:2007aua},
originating from self-dual symmetry of the $AdS_5$ under
T-duality. This dual conformal symmetry can be extended to the
full dual superconformal
symmetry\cite{Berkovits:2008ic,Beisert:2008iq}, considering full
set of supergluon amplitudes. In this case, the existence of
fermionic T-duality transformation play a key role. The
$AdS_5\times S^5$ Green-Schwarz superstring theory is self-dual
under a combination of bosonic and fermionic T-duality. This fact
explains the superconformal and the dual superconformal invariance
in $\mathcal{N}=4$ SYM.

It turns out that the T-duality is closely related to the
integrability of the sigma models\cite{Ricci:2007eq,Beisert:2008iq}.
For $AdS_n$ background, the sigma models are self-dual under bosonic
T-duality, suggesting the local Noether charges of dual model are
related to the non-local charges of the original model and vice
versa. This relation could be generalized to the integrable
super-coset models which are self-dual under the combination of
bosonic and fermionic T-duality transformations. In fact, both
bosonic and fermionic T-duality could be understood as the discrete
automorphism of the global symmetry algebra.


It is interesting to investigate if and under what conditions
other integrable sigma models could be self-dual under fermionic
T-duality. In \cite{Adam:2009}, the authors considered more
general integrable Green-Schwarz sigma models on AdS backgrounds.
They showed that the sigma models  on $AdS_p\times S^p(p=2,3)$
background which are supercosets of PSU supergroups are self-dual
under fermionic T-duality, while the non-critical $AdS_2$ and
$AdS_4$ models and the critical $AdS_4\times CP^3$ which all are
supercosets of OSp supergroups are not. They also argued that in
general the models which are supercosets of ortho-symplectic
groups are not self-dual under fermionc T-duality, hence are short
of the dual superconformal symmetry even its dual models exist. It
was argued that the absence of fermionic self-duality in OSp modes
is due to the lack of appropriate fermionic quadratic terms,
because the Cartan-Killing bilinear form of OSp group is only
nonvanishing for the products of different fermionic generators.

In this paper, we explore this problem further by analyzing other
integrable Green-Schwarz sigma models on AdS backgrounds. We
consider integrable supercosets with $Z_4$ grading.  The existence
of $Z_4$ grading of supercosets allows us to construct
one-parameter families of flat currents\cite{Bena:2003wd,Bin
chen:2005}, which in turn allow for the construction of infinitely
many non-local charges\cite{Luscher:1977rq}. We first show that
the sigma model with $AdS_5\times S^1$ background, which is the
supercoset of $SU(2,2|2)$ supergroup, is self-dual under fermionic
T-duality. We then present a series of new integrable
Green-Schwarz sigma models with the backgrounds $AdS_2\times
CP^n$. Considering the critical dimension of superstring, we only
focus on the cases with $n\leq 4$. These backgrounds could be
taken as supercosets of SU supergroups for arbitrary $n$. However,
for $n=1,3$ the backgrounds could also be realized as supercosets
of OSp supergroups. We show explicitly that all of the SU cases
are self-dual under fermionic T-duality, while the OSp cases are
not. Our study on $AdS_2\times CP^n$ with $n=1,3$ shows that even
the bosonic background is the same, the different
supersymmetrizations may have different behavior under fermionic
T-duality.

Moreover, we find that in the $n=1$ Osp supergroup case,
corresponding to $AdS_2\times CP^1$ background, even though the
sigma model has regular fermionic quadratic term, it fails to be
self-dual under fermionic T-duality. The failure is due to the
shortage of  $\kappa$-symmetry to gauge away the right number of
fermionic degrees of freedom. This happens for other backgrounds,
 including $AdS_2\times S^{2n}$ and
$AdS_4\times S^{2n}$.  Therefore, in general, the sigma models on
supercosets of OSp supergroup can not be self-dual under fermionic
T-duality, but due to different reasons. For OSp supergroups with
superalgebra of types $C(n)$ and $D(m,n)$, the failure stems from
the singular fermionic quadratic terms, while for OSp supergroups
with superalgebra of type $B(m,n)$, the failure comes from the
shortage of $\kappa$-symmetry.


This paper is organized as follows. In Section 2 we show that the
$AdS_5\times S^1$ background  is self-dual under a combination of
bosonic and fermionic T-duality. In section 3, we study the
$AdS_2\times CP^n$ cases. We first discuss the SU cases. After
presenting their superalgebra and $Z_4$ automorphism which are
crucial for the integrability,  we preform the T-duality via
Buscher's procedure, and show that the supercoset models are
self-dual under T-duality. Then we turn to the OSp cases, and show
that they are not self-dual, due to different reasons. In section
4, we conclude and present a brief discussion. We collect some
technical details into the appendices. In appendix A, we give the
definition of the generators of the superalgebra $SU(1,1|n)$. And
in appendix B, we discuss the $\kappa$-symmetry in $AdS_{2n}\times
S^{2m}$ backgrounds.

\section{$AdS_5\times S^1$ background}

In this section we consider the Green-Schwarz sigma-model on
$AdS_5\times S^1$ using the supercoset manifold
$SU(2,2|2)/(SO(4,1)\times SO(3))$. It was first pointed out by
Polyakov \cite{Polyakov:2004br} that noncritical $AdS_p \times
S^q$ are conformal invariant and should be dual to gauge theories
with less or no supersymmetries. And later on in
\cite{Klebanov:2004ya}, Klebanov and Maldacena found the $AdS_5
\times S^1$ solution in the low energy supergravity effective
action of six dimensional noncritical string theory with
Ramond-Ramond flux and in the presence of space-time filling
D5-branes. This solution has the right structure to be dual to
${\cal N}=1$ supersymmetric gauge theories with flavors, in
agreement with the proposal in \cite{Polyakov:2004br}. It has been
shown that such background could be realized as integrable
supercoset with $Z_4$ structure\cite{Bin chen:2005}.

For $AdS_5\times S^1$ background, the $su(2,2|2)$ algebra and its
$Z_4$ structure were studied in \cite{Bin chen:2005}. Here we
redefine the generator as
\begin{eqnarray}
&&D=M_{45}, \quad  P_a=M_{a5}-M_{a4}, \quad
K_a=M_{a5}+M_{a4}, \nonumber\\
&&Q^{\alpha\alpha'}=\frac{1}{2}\varepsilon^{\alpha\beta}C^{\alpha'\beta'}(Q^1_{\beta\beta'}-iQ^2_{\beta\beta'}),
\quad
\bar{Q}^{\dot{\alpha}}_{\alpha'}=-\frac{1}{2}(Q^{1\dot{\alpha}}_{\alpha'}+iQ^{2\dot{\alpha}}_{\alpha'}),
\nonumber \\
&&S_{\alpha\alpha'}=-\frac{1}{2}(Q^1_{\alpha\alpha'}+iQ^2_{\alpha\alpha'}),
\quad
\bar{S}_{\dot{\alpha}}^{\alpha'}=\frac{1}{2}\varepsilon_{\dot{\alpha}\dot{\beta}}C^{\alpha'\beta'}(Q^{1\dot{\beta}}_{\beta'}-iQ^{2\dot{\beta}}_{\beta'}),
\end{eqnarray}
then we have the non-trivial brackets of the algebra
\begin{eqnarray}
&&[D,P_a]=P_a,\quad  [D,K_a]=-K_a, \quad  [P_a,K_b]=2\eta_{ab}D+2M_{ab},\nonumber \\
&&[P_a,M_{bc}]=\eta_{ab}P_c-\eta_{ac}P_b, \quad
[K_a,M_{bc}]=\eta_{ab}K_c-\eta_{ac}K_b, \nonumber\\
&&[M_{ab},M_{cd}]=\eta_{ad}M_{bc}+\eta_{bc}M_{ad}-\eta_{ac}M_{bd}-\eta_{bd}M_{ac}, \nonumber\\
&&[D,Q^{\alpha\alpha'}]=\frac{1}{2}Q^{\alpha\alpha'}, \quad
[D,S_{\alpha\alpha'}]=-\frac{1}{2}S_{\alpha\alpha'},\quad [T_{a'},T_{b'}]=\varepsilon_{a'b'c'}T_{c'},\nonumber\\
&&[P_a,S_{\alpha\alpha'}]=-i\bar{Q}^{\dot{\alpha}}_{\alpha'}(\bar{\sigma}_a)_{\dot{\alpha}\alpha},
\quad
[K_a,Q^{\alpha\alpha'}]=-i\bar{S}^{\dot{\alpha}\alpha'}(\bar{\sigma}_a)_{\dot{\alpha}\alpha},
\nonumber \\
&&[M_{ab},Q^{\alpha\alpha'}]=\frac{1}{2}Q^{\beta\alpha'}(\sigma_{a\bar{b}})^{\alpha}_{\beta},\quad
[M_{ab},S_{\alpha\alpha'}]=\frac{1}{2}S_{\beta\alpha'}(\sigma_{\bar{a}b})^{\alpha}_{\beta},
\nonumber \\
&&[T_{a'},Q^{\alpha\alpha'}]=\frac{1}{2}Q^{\alpha\beta'}(\tau_{a'})^{\beta'}_{\alpha'},
\quad
[T_{a'},S^{\alpha\alpha'}]=\frac{1}{2}S^{\alpha\beta'}(\tau_{a'})^{\beta'}_{\alpha'}, \nonumber \\
&&[Q^{\alpha\alpha'},R]=\frac{i}{2}Q^{\alpha\alpha'},\quad
[S_{\alpha\alpha'},R]=-\frac{i}{2}S_{\alpha\alpha'},\nonumber \\
&&\{Q^{\alpha\alpha'},\bar{Q}^{\dot{\alpha}}_{\beta'}\}=(\sigma^a)^{\alpha\dot{\alpha}}\delta^{\alpha'}_{\beta'}P_a,
\quad
\{S_{\alpha\alpha'},\bar{S}_{\dot{\alpha}}^{\beta'}\}=(\sigma^a)_{\alpha\dot{\alpha}}\delta_{\alpha'}^{\beta'}K_a,
\nonumber \\
&&\{Q^{\alpha\alpha'},S_{\beta\beta'}\}=\delta^{\alpha}_{\beta}\delta^{\alpha'}_{\beta'}\Big
[
i(D+\frac{1}{2}\gamma^{ab}M_{ab})-R\Big]-2i\delta^{\alpha}_{\beta}(\tau^{a'})^{\alpha'}_{\beta'}T_{a'}.
\end{eqnarray}
Here $a,b=0,1,2,3$ are the $so(1,3)$ indices, $\alpha,\beta=1,2$
and $\dot{\alpha},\dot{\beta}=1,2$ are the $so(1,3) $ spinor
indices, which are lowered and raised using
$\epsilon_{12}=-\epsilon_{21}=1$,
$\epsilon^{12}=-\epsilon^{21}=-1$,
$\epsilon_{\dot{1}\dot{2}}=-\epsilon_{\dot{2}\dot{1}}=1$,$\epsilon^{\dot{1}\dot{2}}=-\epsilon^{\dot{2}\dot{1}}=-1$
. The matrices $(\eta_{ab})=(\eta^{ab})=\mathrm{diag}(-+++)$, and
the Dirac matrices are formed by $\sigma^a=(\mathbf{1},\sigma^i)$,
$\bar{\sigma}^a=(\mathbf{1},-\sigma^i)$, $\sigma^{a\bar
b}=\frac{1}{2}[\sigma^a,\bar{\sigma}^b]$. And $a',b'=1,2,3$ are
the $so(3)$ indices, $\alpha',\beta'=1,2$ are the $so(3)$ spinor
indices, which are lowered and raised using
$C_{\alpha'\beta'}=\eta_{\alpha'\beta'}$. The matrices
$(\eta_{a'b'})=\mathrm{diag}(+++)$,  and the Dirac matrices are
$\tau_{a'}=-\mathrm{i}\sigma_{a'}$. $T_{a'}$ and $R$ are the
generators of $su(2)$ and $u(1)$ respectively.

The $Z_4$-automorphism invariant subspaces are classified as
\begin{eqnarray}
\mathcal{H}_0 & = & \{ P_a-K_a, J_{ab},T_{a'}\}, \ \nonumber \\
\mathcal{H}_1 & = & \{ \varepsilon_{\alpha\beta}C_{\alpha'\beta'}Q^{\beta\beta'}-S_{\alpha\alpha'}, \varepsilon^{\dot{\alpha}\dot{\beta}}C_{\alpha'\beta'}\bar{S}_{\dot{\beta}}^{\beta'}-\bar{Q}^{\dot{\alpha}}_{\alpha'} \}, \ \nonumber \\
\mathcal{H}_2 & = & \{ P_a+K_a, D, R \}, \ \nonumber \\
\mathcal{H}_3 & = & \{
\varepsilon_{\alpha\beta}C_{\alpha'\beta'}Q^{\beta\beta'}+S_{\alpha\alpha'},\varepsilon^{\dot{\alpha}\dot{\beta}}C_{\alpha'\beta'}\bar{S}_{\dot{\beta}}^{\beta'}+\bar{Q}^{\dot{\alpha}}_{\alpha'},
\}
\end{eqnarray}
where $\mathcal{H}_i$ denotes the subspace of grading $i$.

The non-vanishing components of the Cartan-Killing bilinear forms
are
\begin{eqnarray}
\mathrm{Str}(P_aK_b)=-2\eta_{ab},\quad \mathrm{Str}(DD)=1,\quad
\mathrm{Str}(J_{ab}J_{cd})=\eta_{ac}\eta_{bd}-\eta_{ad}\eta_{bc},
\nonumber \\
\mathrm{Str}(RR)=4,\quad
\mathrm{Str}(T_{a'}T_{b'})=-\frac{1}{2}\delta_{a'b'}, \quad
\mathrm{Str}(Q^{\alpha\alpha'}S_{\beta\beta'})=2i\delta^{\alpha}_{\beta}\delta^{\alpha'}_{\beta'}.
\end{eqnarray}

A general group element $g\in SU(2,2|2)$ can be parameterized as
\begin{eqnarray}
g=\mathrm{exp}(x^aP_a+x'^aK_a+\theta_{\alpha\alpha'}Q^{\alpha\alpha'}+\xi^{\alpha\alpha'}S_{\alpha\alpha'})\mathrm{exp}(\bar{\theta}_{\dot{\alpha}}^{\alpha'}\bar{Q}^{\dot{\alpha}}_{\alpha'}+\bar{\xi}^{\dot{\alpha}}_{\alpha'}\
\bar{S}^{\alpha'}_{\dot{\alpha}})y^D\mathrm{exp}(R).
\end{eqnarray}
Now we use the $\kappa$-symmetry to fix $\xi^{\alpha\alpha'}=0$,
and use the gauge symmetry to set $x'^a=0$, then we read the coset
representative
\begin{eqnarray}
g&=&\mathrm{exp}(x^aP_a+\theta_{\alpha\alpha'}Q^{\alpha\alpha'})\mathrm{exp}(\bar{\theta}_{\dot{\alpha}}^{\alpha'}\bar{Q}^{\dot{\alpha}}_{\alpha'}+\bar{\xi}^{\dot{\alpha}}_{\alpha'}\
\bar{S}^{\alpha'}_{\dot{\alpha}})y^D\mathrm{exp}(R), \nonumber
\\
&\equiv&\exp(x^a P_a+\theta_{\alpha\alpha'}Q^{\alpha\alpha'})e^B.
\end{eqnarray}

The Green-Schwarz sigma-model on the supercosets of supergroup $G$
with $\bf {Z}_4$ automorphism is generically described by the
action
\begin{eqnarray}\label{eq:GS-coset-action}
 S = \frac{R^2}{4 \pi \alpha'} \int d^2 z \mathrm{Str} \left( J_2 \bar J_2 +
  \frac{1}{2} J_1 \bar J_3 - \frac{1}{2} J_3 \bar J_1\right) \ ,
\end{eqnarray}
where R is the $AdS$ radius, $J = g^{-1} \partial g$ for $g \in G$
and $J_i$ is the current $J$ restricted to the invariant subspace
$\mathcal{H}_i$ of the $\bf {Z}_4$ automorphism of the algebra of
the group $G$. In the case at hand, using the above algebra, the
sigma-model (\ref{eq:GS-coset-action}) takes the form
\begin{eqnarray}\label{eq:psu(2,2|2)-GS action}
S&=&\frac{R^2}{4\pi\alpha'}\int d^2z \bigg[-(J_{P_a}+J_{K_a})(\bar
J_{P_b}+\bar J_{K_b})\eta^{ab} +J_D\bar J_D+4J_R\bar J_R \nonumber\\
&&+i\varepsilon^{\alpha\beta}C^{\alpha'\beta'}(J_{Q_{\alpha\alpha'}}\bar{J}_{Q_{\beta\beta'}}-J_{S_{\alpha\alpha'}}\bar{J}_{S_{beta\beta'}})+
i\varepsilon_{\dot{\alpha}\dot{\beta}}C^{\alpha'\beta'}(J_{\bar{Q}^{\dot{\alpha}}_{\alpha'}}\bar{J}_{\bar{Q}^{\dot{\beta}}_{\beta'}}-J_{\bar{S}^{\dot{\alpha}}_{\alpha'}}\bar{J}_{\bar{S}^{\dot{\beta}}_{\beta'}})\bigg],
\end{eqnarray}
where the currents take the form
\begin{eqnarray}
  J_{P_a}& = & [e^{-B} (dx^a P_a + d\theta_{\alpha\alpha'}Q^{\alpha\alpha'}) e^B]_{P_a} \ , \quad
  J_{Q^{\alpha\alpha'}} = [e^{-B} (dx^a P_a + d\theta_{\alpha\alpha'}Q^{\alpha\alpha'}) e^B]_{Q^{\alpha\alpha'}} \ , \nonumber \\
  J_K & = & 0 \ , \quad
  J_{\bar{Q}^{\dot{\alpha}}_{\alpha'}} = [e^{-B} de^B]_{\bar{Q}^{\dot{\alpha}}_{\alpha'}} \ , \quad
  J_{S_{\alpha\alpha'}} = 0 \ , \quad
  J_{\bar {S}_{\dot{\alpha}}^{\alpha'}} = [e^{-B} de^B]_{\bar {S}_{\dot{\alpha}}^{\alpha'}} \ , \nonumber \\
  J_D & = & [e^{-B} de^B]_D \ , \quad
  J_{R} = [e^{-B} de^B]_{R} \ .
\end{eqnarray}

We can now T-dualize the action with respect to $x^a$ and
$\theta_{\alpha\alpha'}$ via Buscher's
procedure\cite{Buscher:1987qj}. By introducing the bosonic gauge
fields $(A^a,\bar A^a)$ for the translation $P_a$, the fermionic
gauge fields $(A_{\alpha\alpha'},\bar A_{\alpha\alpha'})$ for the
supercharges $Q^{\alpha\alpha'}$, and
 the Lagrange multipliers $\tilde{x}_a$ and
 $\tilde{\theta}^{\alpha\alpha'}$, adding the Lagrange multiplier
 term
\begin{eqnarray}
\frac{R^2}{4\pi \alpha'}\int
d^2z[\tilde{x_a}(\bar{\partial}A^a-\partial \bar
A^a)+\tilde{\theta}^{\alpha\alpha'}(\bar{\partial}A_{\alpha\alpha'}-\partial
\bar A_{\alpha\alpha'})]
\end{eqnarray}
to the action (\ref{eq:psu(2,2|2)-GS action}), we have  the full
action
\begin{eqnarray}
   S & = & \frac{R^2}{4 \pi \alpha'} \int d^2 z [ -\eta_{ab} A'^a
    \bar A'^b + i\varepsilon^{\alpha\beta}C^{\alpha'\beta'} A'_{\alpha\alpha'} \bar
    A'_{\beta\beta'}
    + \dots \nonumber \\
    && +\tilde{x_a}(\bar{\partial}A^a-\partial \bar
A^a)+\tilde{\theta}^{\alpha\alpha'}(\bar{\partial}A_{\alpha\alpha'}-\partial
\bar A_{\alpha\alpha'})]
\end{eqnarray}
where \dots denotes the spectator terms and
\begin{eqnarray}
A'^a=[e^{-B} (A^b P_b + A_{\alpha\alpha'}Q^{\alpha\alpha'})
e^B]_{P_a},\quad A'_{\alpha\alpha'}=[e^{-B} (A^a P_a +
A_{\beta\beta'}Q^{\beta\beta'})_e^B]_{Q^{\alpha\alpha'}}.
\end{eqnarray}
After plugging the inverse relations $A^a=[e^{B} (A'^b P_b +
A'_{\alpha\alpha'}Q^{\alpha\alpha'}) e^{-B}]_{P_a}$ and
$A_{\alpha\alpha'}=[e^{B} (A'^a P_a +
A'_{\beta\beta'}Q^{\beta\beta'})e^{-B}]_{Q^{\alpha\alpha'}}$ into
the action, we can integrate out $A'^a$ and $A'_{\alpha\alpha'}$
by using their equations of motion
\begin{eqnarray}
  A'^a & = & \eta^{ab}([e^B \partial \tilde x_b P_b e^{-B}]_{P_b} + \partial[e^B
    \tilde \theta^{\alpha\alpha'} P_b e^{-B}]_{Q^{\alpha\alpha'}}) =  \eta^{ab}[e^{-B} (\partial
    \tilde x_c K_c + i \partial \tilde \theta^{\alpha\alpha'} S_{\alpha\alpha'}) e^B]_{K_b} \ ,
  \nonumber \\
  \bar A'^a & = &-\eta^{ab} ([e^B \bar\partial \tilde x_b P_b e^{-B}]_{P_b} + [e^B \bar\partial
    \tilde \theta^{\alpha\alpha'} P_b e^{-B}]_{Q^{\alpha\alpha'}}) = -\eta^{ab}[e^{-B} (\bar\partial
    \tilde x_c K_c + i \bar\partial \tilde \theta^{\alpha\alpha'} S_{\alpha\alpha'}) e^B]_{K_\alpha} \ , \nonumber \\
  A'_{\alpha\alpha'} & = &  -i \varepsilon_{\alpha\beta}C_{\alpha'\beta'} ([e^B \partial \tilde
  x_a
  Q_{\beta\beta'} e^{-B}]_{P_a} - [e^B \partial \tilde \theta^{\gamma\gamma'} Q_{\beta\beta'}
    e^{-B}]_{Q_{\gamma\gamma'}}) \nonumber\\
  & = &  \varepsilon_{\alpha\beta}C_{\alpha'\beta'} [e^{-B} (\partial \tilde x_a K_a + i \partial
    \tilde \theta^{\gamma\gamma'} S_{\gamma\gamma'}) e^B]_{S_{\beta\beta'}}  , \nonumber \\
  \bar  A'_{\alpha\alpha'} & = &  i \varepsilon_{\alpha\beta}C_{\alpha'\beta'} ([e^B \bar\partial \tilde
  x_a
  Q_{\beta\beta'} e^{-B}]_{P_a} - [e^B \bar\partial \tilde \theta^{\gamma\gamma'}
  Q_{B\beta'}
  e^{-B}]_{Q_{\gamma\gamma'}}) \nonumber\\
  & = & -\varepsilon_{\alpha\beta}C_{\alpha'\beta'} [e^{-B} (\bar\partial \tilde x_a K_a + i \bar\partial
  \tilde \theta^{\gamma\gamma'} S_{\gamma\gamma'})
  e^B]_{S_{\beta\beta'}}.
\end{eqnarray}
Finally we obtain the T-dualized action
\begin{eqnarray}\label{eq:psu(2,2|2)-T-dual action}
 S&=&\frac{R^2}{4\pi\alpha'}\int d^2z \bigg[-[e^{-B} (\partial
    \tilde x_c K_c + i \partial \tilde \theta^{\alpha\alpha'} S_{\alpha\alpha'})
    e^B]_{K_a}[e^{-B} (\partial
    \tilde x_c K_c + i \partial \tilde \theta^{\alpha\alpha'} S_{\alpha\alpha'})
    e^B]_{K_b}\eta^{ab} \nonumber\\
    {}&&-i\varepsilon^{\alpha\beta}C^{\alpha'\beta'}[e^{-B} (\bar\partial \tilde x_a K_a + i \bar\partial
    \tilde \theta^{\gamma\gamma'} S_{\gamma\gamma'}) e^B]_{S_{\alpha\alpha'}}[e^{-B} (\bar\partial \tilde x_a K_a + i \bar\partial
    \tilde \theta^{\gamma\gamma'} S_{\gamma\gamma'})
    e^B]_{S_{\beta\beta'}}\nonumber\\
    {}& &+\dots\bigg]
\end{eqnarray}
Note that the $su(2,2|2)$ algebra admits the automorphism
\begin{eqnarray}
 P_a\leftrightarrow K_a \ ,
  \quad D \to -D \ ,
  \quad  Q^{\alpha\alpha'} \leftrightarrow S_{\alpha\alpha'} \ , \quad
  \bar Q^{\dot{\alpha}}_{\alpha'} \leftrightarrow \bar S_{\dot{\alpha}}^{\alpha'} \ ,
\end{eqnarray}
with the rest of the generators unchanged. Applying this
automorphism combined with the change of variables
\begin{eqnarray}
 x \to \tilde x \ , \quad
  \theta_{\alpha\alpha'} \to i \tilde \theta^{\alpha\alpha'}\ , \quad
  \bar \theta_{\dot{\alpha}}^{\alpha'} \leftrightarrow \bar \xi^{\dot{\alpha}}_{\alpha'} \ , \quad
  y \to y \ ,
\end{eqnarray}
to (\ref{eq:psu(2,2|2)-GS action}),  we recover
(\ref{eq:psu(2,2|2)-T-dual action}). This completes our proof that
the background $AdS_5\times S^1$ is self-dual under fermionic
T-duality.

\section{$AdS_2\times CP^n$ background}

In this section, we turn to the sigma models on the $AdS_2\times
CP^n$ backgrounds. We restrict ourselves to the critical and
noncritical superstrings with $n\neq 4$. For $n=1$, since $CP^1$
is just two-dimensional sphere $S^2$, we have $AdS_2\times S^2$,
which has been studied in \cite{Adam:2009}. The superstring
propagating in the $\mathrm{AdS_2\times CP^n}$ background has the
bosonic part
\begin{equation}
\mathrm{AdS_2\times CP^n\cong SO(1,2)/U(1)\times SU(n+1)/U(n)}.
\end{equation}
The supergroups which have bosonic subgroups
$\mathrm{SO(1,2)\times SU(n+1)}$ can be $\mathrm{SU(1,1|n+1)}$ for
$n=1,2,3,4$, and $\mathrm{OSp(3|2)}$ for $n=1$ and
$\mathrm{OSp(6|2)}$ for $n=3$. This means that for $n=1,3$, we may
have two different supercosets realizations, based on SU
supergroups or on OSp supergroups, with different supercharges
respectively.

\subsection{PSU supergroup case}

In this subsection, we will focus on the $\mathrm{SU(1,1|n+1)}$
case, with a sigma-model on the coset space
\begin{eqnarray}
\mathrm{\frac{SU(1,1|n+1)}{U(1)\times U(n)\times U(1)}}
\end{eqnarray}
The last $U(1)$ is the overall generator. The super-Lie algebras
$su(1,1|n)$ are the algebras of $(2+n)\times(2+n)$ matrices with
bosonic diagonal blocks and fermionic off-diagonal blocks
\begin{eqnarray}
M=\pmatrix{ A & X \cr  Y & B} ~~~{\rm with}~~tr A = tr B = 0,
\end{eqnarray}
where $A$ and $B$ are even(bosonic) $2\times 2$ and $n\times n$
matrices. The $2\times n$ matrix $X$ and $n\times 2$ matrix $Y$
are odd. The anti-hermiticity condition is
\begin{eqnarray}
M^\dagger \equiv \pmatrix{ \sigma_3A^\dagger \sigma_3 & -i\sigma_3
Y^\dagger \cr   -iX^\dagger \sigma_3 & B^\dagger} = -M,
\end{eqnarray}
which leads to  \begin{eqnarray} A=-\sigma_3 A^{\dagger} \sigma_3
^{-1} ~,~~B=-B^{\dagger}~,~~X=i \sigma_3 Y^{\dagger}.
\end{eqnarray}
The algebra $su(1,1|n)$ has a ${\bf Z}_4$ automorphism, generated
by the conjugation map $M\to \Omega(M) \equiv {\Omega} M
\Omega^{-1}$ with the matrix
\begin{eqnarray}
\Omega =\pmatrix{ \sigma_3 & 0 & 0 \cr
  0 & i{\bf I}_{n-1} & 0 \cr 0 & 0 & -i}~.
\end{eqnarray}
This conjugation respects the anti-hermiticity conditions given
above and manifestly gives an algebra automorphism, $\Omega^4 (M) =
M$. In addition, the invariant subalgebra $\Omega(M) = M$ is the
desired bosonic $u(1)\oplus u(n-1)\oplus u(1)$ algebra.

Using this automorphism the algebra can be decomposed into ${\bf
Z}_4$-invariant subspaces $\mathcal{H}_k$ ($k = 0 \dots, 3$) such
that
\begin{eqnarray}
  \mathcal{H}_k = \{ X \in \mathrm{su}(1,1|n) | \Omega X \Omega^{-1} =
  i^k X \} \ .
\end{eqnarray}

The $su(1,1|n)$ algebra is generated by the following
(anti)commutators:
\begin{eqnarray}
  &&[D, P]=P , \quad [D, K] = -K , \quad [P, K] = -2 D, \quad
  [R_i^j, R_k^l]=\delta_k^j R_i^l-\delta_i^l R_k^j, \qquad \nonumber   \\
  &&[D, Q^i]=\frac{1}{2} Q^i ,\quad [D, S_i] = -\frac{1}{2} S_i,  \nonumber \\
  &&[P, Q^i]=0 ,\quad [P, S_i] = i\bar Q_i,\quad[K, Q^i]=i\bar S^i ,\quad [K, S_i]=0, \nonumber \\
  &&[R_i^j, Q^k]=-(\delta_i^k Q^j-\frac{1}{n}\delta_i^j Q^k), \quad  [R_i^j, S_k]=(\delta_k^j S_i-\frac{1}{n}\delta_i^j S_k),\ \quad  \nonumber \\
  &&\{ Q^i, \bar Q_j \}=\delta_j^i P ,\quad \{ S_i, \bar S^j \}=-\delta_j^j
  K,\quad \{Q^i, S_j\}=-\mathrm{i}(\delta_j^i (A+D) + R_j^i),\quad  \nonumber \\
  &&[Q^i, A]= -\frac{n-2}{n}Q^i, \quad [S_i, A]= \frac{n-2}{n}S_i,
\end{eqnarray}
where $i,j=1,2,\dots,n$ are $\mathrm{SU(n)}$ R-symmetry indices, and
$A$ is the overall $U(1)$ generator
\begin{eqnarray}
A= \left(
  \begin{array}{cc|cccc}
    1 & 0 & 0 & 0 & \dots & 0\\
    0 & 1 & 0 & 0 &  \dots & 0\\ \hline
    0 & 0 & \frac{2}{n} & 0 & \dots & 0\\
    0 & 0 & 0 & \frac{2}{n} & \dots & 0\\
    \vdots & \vdots & \vdots & \vdots & \ddots & \vdots \\
    0 & 0 & 0 & 0 & \dots & \frac{2}{n}
  \end{array}
  \right) \ .
\end{eqnarray}
The definition of other generators could be found in appendix A.
Notice we neglect the $AdS_2$ spinor index $\alpha=1$ of the
fermionic generators (eg. $Q^{\alpha i}$).

The $\bf{Z}_4$-graded subspaces of the algebra are
\begin{eqnarray}
\mathcal{H}_0 & = & \{ P+K, R_a^b, A \}, \ \nonumber \\
\mathcal{H}_1 & = & \{ Q^a-\bar S^a, Q^d+\bar S^d, \bar Q_a-S_a, \bar Q_d+S_d \}, \ \nonumber \\
\mathcal{H}_2 & = & \{ P-K, D, R_n^a,R_a^n \}, \ \nonumber \\
\mathcal{H}_3 & = & \{ Q^a+\bar S^a, Q^d-\bar S^d, \bar Q_a+S_a,
\bar Q_d-S_d \},
\end{eqnarray}
where $a,b=1,2,\cdots ,n-1$. The non-vanishing components of the
Cartan-Killing bilinear form are
\begin{eqnarray}
\mathrm{Str}(PK) =-1, \quad \mathrm{Str}(DD)=\frac{1}{2}, \ \nonumber \\
\mathrm{Str}(R_i^jR_k^l)=-(\delta_i^l \delta
_k^j-\frac{1}{n}\delta_i^j \delta_k^l),\ \nonumber\\
\mathrm{Str}(Q^iS_j)=-\frac{i}{2}, \quad \mathrm{Str}(\bar Q_i\bar
S^j)=\frac{i}{2}.
\end{eqnarray}

In order to study the fermionic T-duality, it turns out to be
convenient to redefine the generators. Instead of using the above
algebra directly, we redefine the Grassmann-odd generators as
linear combinations of the original ones
\begin{eqnarray}
 \begin{array}{cccccc}
  Q^a & = & \frac{Q^a+\bar Q_n}{2} , & \hat Q_a & = & \frac{Q^a-\bar Q_n}{2}, \nonumber\\
  S^a & = & \frac{\bar S^a-S_n}{2} , & \hat S_a & = & \frac{\bar S^a+S_n}{2}, \ \nonumber\\
  Q^n & = & \frac{\Sigma \bar Q_a-Q_n}{2} ,& \hat Q_n & = & \frac{-Q^n-\Sigma \bar Q_n}{2}, \nonumber\\
  S^n & = & \frac{-\Sigma S_a-\bar S^n}{2} ,& \hat S_a & = & \frac{\Sigma S_a-\bar
  S^n}{2},
 \end{array}
\end{eqnarray}
where the sum is over $1,2,\dots, n-1$. The $\bf {Z}_4$ invariant
subspaces of the algebra change to
\begin{eqnarray}
\mathcal{H}_0 & = & \{ P+K, R_a^b, A \}, \ \nonumber \\
\mathcal{H}_1 & = & \{ Q^a-S^a, Q^n+S^n, \hat Q_a-\hat S_a, \hat Q_n+\hat S_n \}, \ \nonumber \\
\mathcal{H}_2 & = & \{ P-K, D, R_n^a,R_a^n \}, \ \nonumber \\
\mathcal{H}_3 & = & \{ Q^a+S^a, Q^n-S^n, \hat Q_a+\hat S_a, \hat
Q_n-\hat S_n \},
\end{eqnarray}
and the nonvanishing Cartan-Killing bilinear form of the fermionic
generators change to
\begin{eqnarray}
\mathrm{Str}(Q^iS^j)=iC_{ij},\quad \mathrm{Str}(\hat Q_i \hat
S_j)=-iC_{ij},
\end{eqnarray}
where
$$
C_{ij}=\left(
 \begin{array}{ccccc}
 0 & 0 & \dots & 0 & 1\\
 0 & 0 & \dots & 0 & 1\\
 \vdots & \vdots & \ddots & \vdots & \vdots\\
 0 & 0 & \dots & 0 & 1\\
 -1 & -1 & \dots & -1 &0
 \end{array} \right).
$$

The sigma-model action (\ref{eq:GS-coset-action}) now takes the
form
\begin{eqnarray}\label{eq:psu11n-GS action}
  S & = & \frac{R^2}{4 \pi i'} \int d^2 z \Big[ \frac{1}{2} (J_P
    - J_K) (\bar J_P - \bar J_K) + \frac{1}{2} J_D \bar J_D +
    \frac{1}{2} J_{R_a^n} \bar J_{R_n^a}  - \nonumber \\
    && {} - \frac{i}{2} \eta_{i j} (J_{Q_i} \bar
    J_{Q_j} - J_{\hat Q_i} \bar J_{\hat Q_j} +
    J_{S_i} \bar J_{S_j} - J_{\hat S_i} \bar J_{\hat
      S_j}) \Big] \ ,
\end{eqnarray}
where $\eta_{an}=\eta_{na}=1$ and zero otherwise. This is exactly
the same as the one in \cite{Adam:2009} if we take $n=2$.

Next, after fixing the kappa symmetry and the gauge symmetry, we
parameterize the coset element as
\begin{eqnarray}
g= e^{x P + \theta_i Q^i} e^B,
\end{eqnarray}
where
\begin{equation}
e^B\equiv
e^{\hat{\theta}^i\hat{Q}_i+\hat{\xi}^i\hat{S}_i}y^De^{\Sigma
y^i_jR^j_i/y}.
\end{equation}

The components of the Maurer-Cartan 1-form are
\begin{eqnarray}\label{eq:psu11n-Maurer-Cartan-1-form}
  J_P & = & [e^{-B} (dx P + d\theta_i Q^i) e^B]_P \ , \quad
  J_{Q^i} = [e^{-B} (dx P + d\theta_i Q^i)
    e^B]_{Q^i} \ , \nonumber \\
  J_K & = & 0 \ , \quad
  J_{\hat Q_i} = [e^{-B} de^B]_{\hat Q_i} \ , \quad
  J_{S^i} = 0 \ , \quad
  J_{\hat S_i} = [e^{-B} de^B]_{\hat S_i} \ , \nonumber \\
  J_D & = & [e^{-B} de^B]_D \ , \quad
  J_{R_i^j} = [e^{-B} de^B]_{R_i^j} \ .
\end{eqnarray}

We would like to do T-dual transformation in the directions of the
Abelian sub-algebra formed by the generators $P$ and $Q^i$.
Similar to the case in section 2, we introduce the bosonic gauge
fields $(A,\bar A)$ for the translation $P$ and the fermionic
gauge fields $(A_i,\bar A_i)$ for the supercharges $Q^i$, and add
the Lagrange multiplier term
\begin{eqnarray}
\frac{R^2}{4\pi \alpha'}\int d^2z[\tilde{x}(\bar{\partial}A-\partial
\bar A)+\tilde{\theta}^i(\bar{\partial}A_i-\partial \bar A_i)]
\end{eqnarray}
with $\tilde{x}$ and $\tilde{\theta}^i$ being multiplier, to the
action (\ref{eq:psu11n-GS action}). Then the full action takes the
form
\begin{eqnarray}
   S & = & \frac{R^2}{4 \pi \alpha'} \int d^2 z [ \frac{1}{2} A'
    \bar A' - \frac{i}{2} \eta^{ij} A'_i \bar A'_j
    + \dots  \nonumber \\
    && +\tilde{x}(\bar{\partial}A-\partial
\bar A)+\tilde{\theta}^i(\bar{\partial}A_i-\partial \bar A_i)], \
\end{eqnarray}
where
\begin{eqnarray} A' = [e^{-B} (AP + A'_i Q^i) e^B ]_P \ ,
\quad
  A'_i = [e^{-B} (AP + A'_i Q^i) e^B]_{Q^i}.
\end{eqnarray}
With the inverse $A=[e^{B} (dx P + d\theta_i Q^i) e^{-B} ]_P$ and
$A_i=[e^{B} (d P + d\theta_i Q^i) e^{-B}]_{Q^i}$, we find the
equations of motion
\begin{eqnarray}
  A' & = & -2 [e^B \partial \tilde x P e^{-B}]_P - 2[e^B \partial
    \tilde \theta^i P e^{-B}]_{Q^i} = -2 [e^{-B} (\partial
    \tilde x K + i \partial \tilde \theta^i S_i) e^B]_K \ ,
  \nonumber \\
  \bar A' & = & 2 [e^B \bar \partial \tilde x P e^{-B}]_P + 2 [e^B
    \bar \partial \tilde \theta^i P e^{-B}]_{Q^i} =
  2 [e^{-B} (\bar \partial \tilde x K + i \bar \partial \tilde
    \theta^i S_i) e^B]_K \ , \nonumber \\
  A'_i & = & 2 i \eta_{ij} ([e^B \partial \tilde x
    Q^j e^{-B}]_P - [e^B \partial \tilde \theta^k Q^j
    e^{-B}]_{Q^k}) =
  -2 \eta_{ij} [e^{-B} (\partial \tilde x K + i \partial
    \tilde \theta^k S_k) e^B]_{S_j}  , \nonumber \\
  \bar A'_i & = & 2 i \eta_{ij} ([e^B \bar \partial
    \tilde x Q^j e^{-B}]_P - [e^B \bar \partial \tilde
    \theta^k Q^j e^{-B}]_{Q^k}) =
  -2 \eta_{ij} [e^{-B} (\bar \partial \tilde x K + i \bar
    \partial \tilde \theta^k S_k) e^B]_{S_j}
    . \nonumber
\end{eqnarray}
Integrating out $A'$ and $A'^i$, and rescaling $\tilde x \to
\frac{1}{2} \tilde x$ and $\tilde \theta^i \to \frac{1}{2} \tilde
\theta^i$, we have the T-dualized action
\begin{eqnarray}\label{eq:psu11n-T-dual action}
  S_T & = & \frac{R^2}{4 \pi \alpha'} \int d^2 z \Big[ \frac{1}{2}[e^{-B}
      (\partial \tilde x K + i \partial \tilde \theta^i S_i)
      e^B]_K [e^{-B} (\bar \partial \tilde x K + i \bar \partial
      \tilde \theta^i S_i) e^B]_K  \nonumber \\
    && {} - \frac{i}{2} \eta_{ij} [e^{-B} (\partial \tilde x
      K + i \partial \tilde \theta^k S_k) e^B]_{S_i}
    [e^{-B} (\bar \partial \tilde x K + i \bar \partial \tilde
      \theta^k S_k) e^B]_{S_j} + \dots \Big] \ ,
\end{eqnarray}
Using the automorphism of the algebra
\begin{equation}
  P \leftrightarrow K \ ,
  \quad D \to -D \ ,
  \quad  Q^i \leftrightarrow S_i \ , \quad
  \hat Q_i \leftrightarrow \hat S_i \ ,
\end{equation}
and changing the variables
\begin{equation}
  x \to \tilde x \ , \quad
  \theta_i \to i \tilde \theta^i\ , \quad
  \hat \theta_i \leftrightarrow \hat \xi_i \ , \quad
  y_i^j \to \frac{y_i^j}{y^2} \ ,
\end{equation}
we find that the action (\ref{eq:psu11n-GS action}) is the same as
 (\ref{eq:psu11n-T-dual action}). This shows that the supercosets
 of $SU(1,1|n)$ group is self-dual under fermionic T-duality.

\subsection{The $OSp$ case}

In the above subsection we discussed the $AdS_2\times CP^n$
supercoset models based on $PSU$ supergroups. In this section, we
will discuss another realization of $AdS_2\times CP^n$ based on
ortho-symplectic supergroups. As we have said before, there are
only two cases, $OSp(3|2)$ for $n=1$ and $OSp(6|2)$ for $n=3$.

\subsubsection{The $OSp(3|2)$ case}

For n=1 case, there is another supercoset realization of
$AdS_2\times CP^1\cong SO(1,2)/SO(1,1)\times SO(3)/SO(2)$. The
supergroup $OSp(3|2)$ corresponds to the superalgebra $B(1,1)$,
with its bosonic subgroup being $SO(3)\times Sp(2)$. It has six
real fermionic generators transforming as the $(3,2)$
representation of $SO(3)\times Sp(2)$. This is different from its
PSU realization. In these two different realizations, supercharges
are totally different.

The algebra of $osp(3|2)$ is
\begin{eqnarray}
&&[D,P]=P, \quad [D,K]=-K,  \quad [P,K]=-2D, \nonumber \\
&&[D,Q_i]=\frac{1}{2}Q_i,  \ [D,S_i]=\frac{1}{2}S_i \quad [P,S_i]=iQ_i,\quad [K,Q_i]=iS_i, \nonumber \\
&&[J_i,J_j]=i\varepsilon_{ijk}J_k,\quad[J_i,Q_j]=i\varepsilon_{ijk}Q_k,
\quad [J_i,S_j]=i\varepsilon_{ijk}S_k,
\nonumber \\
&&\{Q_i,Q_j\}=\delta_{ij}P, \quad \{S_i,S_j\}=-\delta_{ij}K,\quad
\{ Q_i,S_j \}=-i\delta_{ij}D-\frac{1}{2}\varepsilon_{ijk}J_k,
\end{eqnarray}
where $i=1,2,3$ are the $SO(3)$ indices, and the
$Z_4$-automorphism invariant subspaces are
\begin{eqnarray}
\mathcal{H}_0 & = & \{ P-K, J_2 \}, \ \nonumber \\
\mathcal{H}_1 & = & \{ Q_i+S_i\}, \ \nonumber \\
\mathcal{H}_2 & = & \{ P+K, D, J_1, J_3 \}, \ \nonumber \\
\mathcal{H}_3 & = & \{ Q_i-S_i\}.
\end{eqnarray}
The non-vanishing components of the Cartan-Killing bilinear form
are
\begin{eqnarray}
\mathrm{Str}(PK)=1,\quad \mathrm{Str}(DD)=-\frac{1}{2}, \nonumber\\
\mathrm{Str}(J_iJ_j)=2\delta_{ij}, \quad
\mathrm{Str}(Q_iS_j)=i\delta_{ij}.
\end{eqnarray}

The action take the form
\begin{eqnarray}
   S &=& \frac{R^2}{4 \pi \alpha'} \int d^2 z [ \frac{1}{2}
   (J_P+J_K)(\bar{J}_P+\bar{J}_K)-\frac{1}{2}J_D\bar{J}_D+2J_{J_3}\bar{J}_{J_3}\nonumber \\
   &&+2J_{J_1}\bar{J}_{J_1}-\frac{i}{2}(J_{Q_i}\bar{J}_{Q_i}-J_{S_i}\bar{J}_{S_i})].
\end{eqnarray}

Before performing the T-duality, we would like to discuss a little
bit about the $\kappa$-symmetry. Conventionally, it is expected that
the $\kappa$-symmetry can remove half of the fermionic degrees of
freedom. This is indeed true for supercosets of SU supergroups.
However, for supercoset models of OSp supergroup, this is not the
case any more.  For example,  superstring on $AdS_4\times
CP^3$\cite{Arutyunov:2008,Stefanski:2008ik} has only eight
$\kappa$-symmetry degrees of freedom, rather than the "expected"
number:twelve. Following the procedure in \cite{Arutyunov:2008}, we
find that our sigma model has only two $\kappa$-symmetry degrees of
freedom. The detail is given in appendix B.

After gauge fixing two fermionic parameters $S_1$ and $S_2$, we
have the coset element
\begin{eqnarray}
g=e^{xp+\theta^1Q_1+\theta^2Q_2}e^{\theta^3Q_3+\xi^3S_3}y^De^{\frac{J_iy^i}{y}}.
\end{eqnarray}
Then with the (anti-)communication relation, the components of the
Maurer-Cartan 1-form are
\begin{eqnarray}
J_P=[e^{-B}dxPe^B]_P, \quad J_D=[e^{-B}de^B]_D, \quad
J_{Q_i}=[e^{-B}d\theta^iQ_ie^B]_{Q_i},\nonumber
\\J_{S_3}=[e^{-B}de^B]_{S_{3}},\quad J_{Q_3}=[e^{-B}de^B]_{Q_3}+[e^{-B}dxPe^B]_{Q_3},\nonumber\\
 J_{J_3}=[e^{-B}de^B]_{J_3} \quad
J_{J_1}=[e^{-B}de^B]_{J_1}+[e^{-B}d\theta^2Q_2e^B]_{J_1},
\end{eqnarray}
where $i=1,2$. Using the fact
\begin{eqnarray}
& &[e^{-B}dxPe^B]_P=\frac{1}{y}dx,\quad
[e^{-B}d\theta^iQ_ie^B]_{Q_i}=\frac{1}{y^{1/2}},\nonumber \\
&&[e^{-B}dxPe^B]_{Q_3}=i\xi^3dx, \quad
[e^{-B}d\theta^2Q_2e^B]_{J_1}=\frac{1}{2}d\theta^2\xi^3, \nonumber
\end{eqnarray}
and  $[e^{-B}de^B]_{Q_3}=j_{Q_3}, [e^{-B}de^B]_{J_1}=j_{J_1}$, we
can rewrite the action  as
\begin{eqnarray}
S&=& \frac{R^2}{4 \pi \alpha'} \int d^2 z[
\frac{1}{2}(\frac{\partial
x\bar{\partial}x}{y^2}+\bar{J}_{Q_3}\xi^3\partial
x+J_{Q_3}\xi^3\bar{\partial} x)\nonumber
\\
&&-\frac{i}{2}(\frac{\partial
\theta^i\bar{\partial}\theta^i}{y}-2i\bar{J}_{J_1}\partial
\theta^2\xi^3-2iJ_{J_1}\bar{\partial}\theta^2 \xi^3)+\dots]
\end{eqnarray}
Here we can see that the bosonic part and the fermionic part are
separated, and after the T-duality, we will have  terms
$\bar{J}_{Q_3}\xi^3\partial \hat{x}-J_{Q_3}\xi^3\bar{\partial}
\hat{x}$, and $\bar{J}_{J_1}\partial
\hat{\theta}^2\xi^3-J_{J_1}\bar{\partial}\hat{\theta^2} \xi^3$.
These terms can not be obtained from any automorphism of the
algebra, so the sigma model is not self-dual.

Let us make a few remarks. Firstly,  this sigma model can not be
cataloged to the OSp case discussed in \cite{Adam:2009}. In that
paper,  the sigma models of OSp supergroups discussed belong to
superalgebra $C(n)$ and $D(m,n)$, with fermionic generators
$\{Q,\bar {Q},S,\bar{S}\}$. In this case, the action includes
$$
\eta^{IJ}(J_{Q^I}\bar{J}_{\bar{Q}^J}-J_{\bar{Q}^I}\bar{J}_{Q^J}),
$$
which leads to singular fermionic quadratic terms and can not be
T-dualized. In our case, we only have fermionic generators $\{Q,S\}$
(there are only six fermionic generators), and the quadratic terms
$$
J_{Q^I}\bar{J}_{Q^I}-J_{{S}^I}\bar{J}_{S^I}
$$
in the action. Obviously, the actions with these forms can be
T-dualized. This discussion can easily be generalized to the
$AdS_2\times S^{2n}$ and $AdS_4\times S^{2n}$ backgrounds, with
the supercosets $OSp(2n+1|2)/(SO(2n)\times U(1))$ and
$OSp(2n+1|4)/(SO(2n)\times SO(3,1))$. The algebras of these
supergroups belong to $B(n,1)$ and $B(n,2)$ types, which also have
the fermionic generators $\{Q,S\}$ and the similar regular
fermionic quadratic  terms.

Secondly, this action will only have regular quadratic term and be
self-dual if we can gauge away half of the fermions, i.e. three
superconformal charges $S$'s. But the terms linear in $J$ appear as
there are only two $\kappa$-symmetry degrees of freedom. Such kind
of term forbids the model from being self-dual. The existence of
such term does not depend on the gauge choice. To see this, instead
of choosing $S_1$ and $S_2$, let us gauge away $Q_3$ and $S_3$.
However, even with this more symmetric gauge choice, we still get
terms like
$\bar{J}_D\xi^i\partial\theta^i+J_D\xi^i\bar{\partial}\theta^i$,
which keeps the model from being self-dual. The similar arguments
also apply to $AdS_2\times S^{2n}$ and $AdS_4\times S^{2n}$
backgrounds in which cases $\kappa$-symmetry can only gauged away
two and four fermionic degrees of freedom respectively.

Finally, there is another subtlety regarding the number of
$\kappa$-symmetries in $OSp$ supercoset models. It was observed
 in \cite{Arutyunov:2008} that for the sigma model on $AdS_4\times CP^3$ when the
string moves entirely in $AdS_4$, the $\kappa$--symmetry parameter
$\epsilon$  vanishes so that  the number of $\kappa$-symmetries
increases from eight to twelve, which allows us to gauge away half
of the fermions. However, this does not mean that the model is
self-dual in this case since the model cannot describe the
complete superstring in $AdS_4\times CP^3$ background. Actually,
it was pointed out in \cite{Gomis:2008jt,Grassi:2009yj} that when
the superstring moves entirely in $AdS_4$ the classical
integrability of the model is still an open issue since the model
could not be taken simply as a supercoset anymore. For example,
the string can move in a subspace  which includes $AdS_4$ but is a
twisted superspace rather than a supercoset. In this case, the
usual analysis of integrability of supercoset model could not be
applied and it is not clear if the model is still integrable or
not\cite{Gomis:2008jt}. Nevertheless, if just focused on the
bosonic model, the string in $AdS_4$ is always integrable, as is
well-known. The same issue may happen in the $OSp$ supercoset
models studied in this paper\footnote{We would like to thank D.
Sorokin for clarifications on this issue.} .


\subsubsection{The $OSp(6|2)$ case}

For the n=3 case, the supercoset can be $AdS_2\times CP^3\cong
SO(1,2)/SO(1,1)\times SO(6)/U(3)$. The supergroup $OSp(6|2)$
corresponds to the superalgebra $D(3,1)$, with bosonic subgroup
$SO(6)\times Sp(2)$, and it has twelve real fermionic generators
transforming as in the $(6,2)$ representation of $SO(6)\times
Sp(2)$. It is easy to see that the algebra is similar to the one
of $OSp(6|4)$\cite{Adam:2009}, which corresponds to superalgebra
$D(3,2)$. We can change the $Sp(4)$ generators\cite{Adam:2009}
with $Sp(2)$ generators and neglect the $SO(3,1)$ spinors indices
to get the algebra
\begin{eqnarray}
&&[\lambda_{k\dot l},\lambda_{m\dot n}]=2i(\delta_{m\dot
l}\lambda_{k\dot n}-\delta_{k\dot n}\lambda_{m\dot l}), \quad
[\lambda_{k\dot l},R_{mn}]=2i(\delta_{m\dot l}R_{kn}-\delta_{n\dot l}R_{k m}), \nonumber \\
&&[R_{mn},R_{kl}]=0,\qquad [R_{mn},R_{\dot k\dot l}]=\frac{i}{2}(\delta_{m\dot k}\lambda_{n\dot l}
-\delta_{m\dot l}\lambda_{n\dot k}-\delta_{n\dot k}\lambda_{m\dot l} +\delta_{n\dot l}\lambda_{m\dot k}), \nonumber \\
&&[D,P]=P,\quad [D,K]=-K,\quad [P,K]=-2D, \nonumber \\
&&[D,Q^l]=\frac{1}{2} Q^l, \qquad [D,S^l]=-\frac{1}{2} S^l,
\quad [P,Q^l]=0, \qquad [K,S^l]=0, \nonumber \\
&&[P,S^l]=-iQ^l, \qquad [K,Q^l]=iS^l, \nonumber \\
&&[R_{kl},Q^{\dot p}]=i(\delta^{\dot p l}Q^k-\delta^{\dot p k}Q^l),
\qquad [R_{kl},S^{\dot p}]=-i(\delta^{\dot p l}S^k-\delta^{\dot p k}S^l), \nonumber \\
&&[R_{\dot k\dot l},Q^{p}]=-i(\delta^{p\dot l}Q^{\dot
k}-\delta^{p\dot k}Q^{\dot l}) ,\qquad [R_{\dot k\dot l},S^{p}] =i(\delta^{p\dot l}S^{\dot k}-\delta^{p\dot k}S^{\dot l}), \nonumber \\
&&[\lambda_{k\dot l},Q^{ p}]=2i\delta^{p \dot l}Q^k, \qquad
[\lambda_{k\dot l},S^{ p}]=2i\delta^{p \dot l}S^k, \nonumber \\
&&[\lambda_{k\dot l},Q^{\dot p}]=-2i\delta^{\dot p k}Q^{\dot
l},\qquad [\lambda_{k\dot l},S^{\dot p}]=-2i\delta^{\dot p k}S^{\dot
l}, \nonumber\\
&&\{Q^l,Q^k\}=0, \qquad \{Q^l,Q^{\dot k}\}=-\delta^{l\dot k}P, \nonumber \\
&&\{S^l,S^k\}=0, \qquad \{S^l,S^{\dot k}\}=-\delta^{l\dot k}K, \nonumber \\
&& \{Q^l,S^k\}=-R_{lk}, \qquad \{Q^{\dot l},S^{\dot k}\}=-R_{\dot l\dot k}, \nonumber \\
&&\{Q^l,S^{\dot k}\}=-i\delta^{l\dot k}D+\frac{1}{2}\lambda_{l\dot
k} \quad \{Q^{\dot l},S^{ k}\}=i\delta^{\dot l
k}D+\frac{1}{2}\lambda_{ k \dot l},
\end{eqnarray}
where $k,l=1,2,3$ and the dotted ones are the $\mathrm{3}$ and
$\mathrm{\bar 3}$ of $u(3)$ respectively. Note that
$\lambda_{k\dot l}$'s are the generators of $so(6)$.

The algebra admits the $Z_4$ automorphism and the invariant
subspaces are
\begin{eqnarray}
\mathcal{H}_0 & = & \{ P-K, \lambda_{l\dot{k}} \}, \ \nonumber \\
\mathcal{H}_1 & = & \{ Q^l-S^l, Q^{\dot{l}}-S^{\dot{l}} \}, \ \nonumber \\
\mathcal{H}_2 & = & \{ P+K, D, R_{kl},R_{\dot{k}\dot{l}} \}, \ \nonumber \\
\mathcal{H}_3 & = & \{ Q^l+S^l, Q^{\dot{l}}+S^{\dot{L}} \}.
\end{eqnarray}

Similar to the $OSp(6|4)$ case, it does not have a fermionic
T-duality symmetry because the matrix multiplying the gauge field
is singular.

\section{Conclusion and discussion}

We have shown that the sigma models on $AdS_5\times S^1$ and
$AdS_2\times CP^n$ background realized as supercosets of PSU
supergroups are self-dual under the combination of bosonic and
fermionic T-duality, while the $AdS_2\times CP^n$ background as
the supercosets of OSp $(n=1,3)$ supergroups are not. For $n=3$,
the OSp sigma model is quite similar to the one on $AdS_4 \times
CP^3$, in which case there is no appropriate fermionic quadratic
term to do T-dualization. However, for $n=1$ the $OSp(3|2)$ model
in our case is very different from the OSp case discussed in
\cite{Adam:2009}. This $OSp(3|2)$ model has appropriate fermionic
quadratic term, which allows us to perform fermionic T-duality.
Nevertheless, the model is not self-dual under T-duality, as there
are not enough $\kappa$-symmetry degrees of freedom.

The difference between these $OSp$ cases stems from the fact that
they belong to different superalgebras. The cases in
\cite{Adam:2009} belong to superalgebras $C(n)$ and $D(m,n)$, the
$OSp(3|2)$ case belongs to $B(1,1)$. For the former case, there is
no appropriate fermionic quadratic terms, while for the latter
case, the fermionic quadratic terms are not singular but now
 the $\kappa$-symmetry degrees of freedom are not enough to gauge away
the right number of fermions to allow the model to be self-dual.
This discussion can be generalized to $AdS_2\times S^{2n}$ and
$AdS_4\times S^{2n}$ backgrounds, both of which could be realized
as the supercosets of OSp supergroups with $B(m,n)$ type
superalgebra.

Another lesson from our study is that for some coset models, they
may have different supersymmetrized coset realizations, which have
different behaviors under fermionic T-duality. A typical example
is $AdS_2\times S^2$. This indicates that in the study of these
backgrounds, we need not only care about the bosonic backgrounds ,
but also need to consider the background RR-flux and the
corresponding supersymmetries.

When the superstring moves only in a subspace of the supercoset, the
number of $\kappa$-symmetries may be enhanced. In other words, the
number of  physical fermionic degrees of freedom depends on the
motion of the string. In this case, it would be interesting to study
the classical integrability and the self-dual properties of the
model.

\section*{Acknowledgments}

We would like to thank Z.B. Xu for his participation at the early
stage of the project. BC is grateful to J.B. Wu for valuable
discussions. BC would like to thank KIAS for hospitality during his
visit. The work was partially supported by NSFC Grant
No.10535060,10775002,10975005, NKBRPC (No. 2006CB805905) and RFDP.

\appendix
\section*{A. The definition of the $su(1,1|n)$ generators }
The generators of the algebra can be taken as
\begin{eqnarray}
D&=&\frac{1}{2}\left(
  \begin{array}{cc|c}
  0 & 1 & 0_{n\times 1}\\
  1 & 0 & 0_{n\times 1}\\ \hline
  0_{1\times n} & 0_{1\times n} & 0_{n\times n}
  \end{array} \right),\nonumber \\
P&=&\frac{1}{2}\left(
  \begin{array}{cc|c}
  i & -i & 0_{n\times 1}\\
  1 & 0 & 0_{n\times 1}\\ \hline
  0_{1\times n} & 0_{1\times n} & 0_{n\times n}
  \end{array} \right), \nonumber \\
K&=&\frac{1}{2}\left(
  \begin{array}{cc|c}
  i & i & 0_{n\times 1}\\
  1 & 0 & 0_{n\times 1}\\ \hline
  0_{1\times n} & 0_{1\times n} & 0_{n\times n}
  \end{array} \right), \nonumber\\
R_i^j&=&E_{i+2,j+2}-\delta^i_j\frac{1}{n}\Sigma E_{i+2i+2},\qquad \nonumber \\
Q^i&=&\frac{1}{\sqrt{2}}(E_{1,i+2}+E_{2,i+2}),  \quad \bar
S^i=\frac{1}{\sqrt{2}}(E_{1,i+2}-E_{2,i+2}), \nonumber \\
\bar Q_i&=&\frac{i}{\sqrt{2}}(E_{i+2,1}-E_{i+2,2}), \quad S_i=-\frac{i}{\sqrt{2}}(E_{i+2,1}+E_{i+2,2}), \nonumber \\
\end{eqnarray}
where $i,j=1,2\dots n$, and
 \begin{eqnarray}
 E_{i,j}=\left\{\begin{array}{cc}
 1,& \mbox{at the $i$th line and $j$th row}\\
 0,& \mbox{otherwise.}
 \end{array}\right.
 \end{eqnarray}

\section*{B. $\kappa$-symmetry}

In this section we would like to discuss the $\kappa$-symmetry of
the  sigma models on $AdS_2\times S^{2m}$ ($m= 1,2,3,4$) and
$AdS_4\times S^{2m}$, ($m=1,2,3$) backgrounds. The coset spaces
for these two case are $OSp(2m+1|2)/(SO(2m)\times U(1))$ and
$OSp(2m+1|4)/(SO(2m)\times SO(3,1))$.

The algebra of $osp(2m+1|2n)$ can be realized by supermatrices of the form
\begin{eqnarray}
A=\pmatrix{ X & \theta \cr  \eta &  Y}
\end{eqnarray}
with the condition
$$
X^t=-X, \quad Y^t=-C_{2n}YC^{1}_{2n}, \quad \eta=-C_{2n}\theta^t,
$$
where X and Y are even $(2m+1)\times (2m+1)$ and $2n\times 2n$
matrices respectively, and $\theta$ and $\eta$ are the odd
$(2m+1)\times 2n$ matrix and $2n\times (2m+1)$ matrix
respectively. The matrices $C_{2n}$ for $n=1,2$ are
$$
C_2=\pmatrix{0 & 1 \cr -1 & 0},  \qquad C_4=\pmatrix{0 & 0 & 0 & 1
\cr 0 & 0 & -1 & 0 \cr 0 & 1 & 0 & 0 \cr -1 & 0 & 0 & 0}
$$
To get the $SO(3,1)$ (or $SO(2,1)$) part of $Sp(4)$ (or $Sp(2)$),
an reality condition should be imposed.

These algebras have inner automorphism $\Omega(A)=\Omega A\Omega^{-1}$, where
$$
\Omega=\pmatrix{I_{2m} & 0 & 0 \cr 0 & -1 & 0 \cr 0 & 0 & \sigma_1}
$$
for $n=1$ and
$$
\Omega=\pmatrix{I_{2m} & 0 & 0 \cr 0 & -1 & 0 \cr 0 & 0 & C_4}
$$
for $n=2$. The $Z_4$-graded subspaces are defined by
$$
 \mathcal{H}_k = \{ X \in \mathrm{osp}(2m+1|2n) | \Omega X \Omega^{-1} =  i^k X
 \}.
$$

The cosets $AdS_{2n}\times S^{2m}$ can be parameterized by the
generators belonging to $\mathcal{H}_2$. Thus a Lie algebra
element parameterizing these cosets can be presented in the form
$$
A=\pmatrix{y_iT_i & 0 \cr 0 & x_\mu T^\mu}
$$
 where $T_i=E_{i,2m+1}-E_{2m+1,i}, i=1,2,\dots,2m$,
$$
T^0=\pmatrix{1 & 0 \cr 0 & -1}, \quad T^1=\pmatrix{0 & 1 \cr -1 & 0}
$$
for $n=1$ and
$$
T^0=\pmatrix{1 & 0 & 0 & 0 \cr 0 & 1 & 0 & 0 \cr 0 & 0 & -1 & 0 \cr
0 & 0 & 0 & -1}, \quad T^1=\pmatrix{0 & 0 & 1 & 0 \cr 0 & 0 & 0 & -1
\cr -1 & 0 & 0 & 0 \cr 0 & 1 & 0 & 0}
$$
$$
T^2=\pmatrix{0 & 0 & 0 & 1 \cr 0 & 0 & 1 & 0 \cr 0 & -1 & 0 & 0 \cr
-1 & 0 & 0 & 0}, \quad T^1=\pmatrix{0 & 0 & 0 & -i \cr 0 & 0 & i & 0
\cr 0 & i & 0 & 0 \cr -i & 0 & 0 & 0}
$$
for $n=2$.

As proved in \cite{Arutyunov:2008}, the $\kappa$-symmetry of the
coset models can be understood as the local fermionic symmetry
with transformation parameters $\varepsilon^{(1)}$ and
$\varepsilon^{(3)}$. $\varepsilon^{(1)}$ takes the form
\begin{eqnarray}\label{eq:definition of epsilo}
\varepsilon^{(1)}=A^{(2)}_{\alpha,-} A^{(2)}_{\beta,-}
\kappa^{\alpha\beta}_{++}+A^{(2)}_{\alpha,-}
\kappa^{\alpha\beta}_{++}A^{(2)}_{\beta,-}
+\kappa^{\alpha\beta}_{++} A^{(2)}_{\alpha,-} A^{(2)}_{\beta,-}
-\frac{1}{2n}\mathrm{str}(\Sigma A^{(2)}_{\alpha,-}
A^{(2)}_{\beta,-}) \kappa^{\alpha\beta}_{++},
\end{eqnarray}
where $\alpha$, $\beta$ are the world sheet indices,
$A^{(2)}_\alpha$ is the current restrict to $\mathcal{H}_2$,
$\kappa^{\alpha\beta}\in \mathcal{H}_1$ is the $\kappa$-symmetry
parameter which is assumed to be independent on the dynamical
fields of these models. The subscript $\pm$ in the above relation
denotes the components are defined with respect to the projections
defined by
$V^\alpha_\pm=\frac{1}{2}(\gamma^{\alpha\beta}\pm\epsilon^{\alpha\beta})V_{\beta}$
with $\gamma^{\alpha\beta}$ being the Weyl-invariant world-sheet
metric. $\varepsilon^{(3)}$ takes a similar form. It is essential
that the action remains invariant under these transformations
without using the equations of motion. Thus the $\kappa$-symmetry
degrees of freedom depend on the rank of $\varepsilon$.

Without loss of generality we can assume that the transversal
fluctuations are all suppressed and the corresponding element
$A^{(2)}$ has the form
$$
A^{(2)}=\pmatrix{yT_0 & 0 \cr 0 & ixT^0}
$$
where $T^0$ corresponds to time direction in the AdS Space and any
element from the tangent space to $S^{2m}$ can be brought to $T_0$
by $SO(2m)$ transformation. Notice that the Virasoro constraint
$\mathrm{str}(A^{(2)}_{\alpha,-}A^{(2)}_{\beta,-})=0$ demands
$x^2=y^2$ for $n=1$ and $2x^2=y^2$ for $n=2$. Plugging this
together with a generic parameter $\kappa$ into
eq.(\ref{eq:definition of epsilo}), we find that the $\varepsilon$
depends on only 2 (for $n=1$) or 4 (for $n=2$) independent complex
fermionic parameters. The reality condition reduces this number by
half. Thus, the $\kappa$-symmetry transformation depends on 2 or 4
real fermions. Consequently the same number of fermionic degrees
of freedom can be gauged away.

We have used the above method to discuss the $\kappa$-symmetry of
the supercosets of SU supergroups and recovered the well-known
result in thess cases successfully.

\end{document}